\documentclass[aps,prb,onecolumn,showpacs,superscriptaddress,amsmath,amssymb]{revtex4-1}      

\usepackage{graphicx}
\usepackage{dcolumn}
\usepackage{bm}
\usepackage{amsmath}
\usepackage{amssymb}
\usepackage{graphicx}
\usepackage{indentfirst}
\usepackage{booktabs}
\usepackage{multirow}
\usepackage{colortbl}
\usepackage{color}

\selectfont

\begin{document}
\title{Quantum Anomalous Hall and Valley Quantum Anomalous Hall Effects in Two-Dimensional d$^0$ Orbital XY Monolayers}

\author{Kang Wang}
\address{State Key Laboratory for Mechanical Behavior of Materials, Center for Spintronics and Quantum System, School of Materials Science and Engineering, Xi'an Jiaotong University, Xi'an, Shaanxi, 710049, China}
\author{Yihui Li}
\address{State Key Laboratory for Mechanical Behavior of Materials, Center for Spintronics and Quantum System, School of Materials Science and Engineering, Xi'an Jiaotong University, Xi'an, Shaanxi, 710049, China}
\author{ Haoliang Mei}
\address{State Key Laboratory for Mechanical Behavior of Materials, Center for Spintronics and Quantum System, School of Materials Science and Engineering, Xi'an Jiaotong University, Xi'an, Shaanxi, 710049, China}
\author{Ping Li}
\email{pli@xjtu.edu.cn}
\address{State Key Laboratory for Mechanical Behavior of Materials, Center for Spintronics and Quantum System, School of Materials Science and Engineering, Xi'an Jiaotong University, Xi'an, Shaanxi, 710049, China}
\address{Key Laboratory for Computational Physical Sciences (Ministry of Eduction), Fudan University, Shanghai, 200433, China}
\address{State Key Laboratory for Surface Physics and Department of Physics, Fudan University, Shanghai, 200433, China}
\author{Zhi-Xin Guo}
\email{zxguo08@xjtu.edu.cn}
\address{State Key Laboratory for Mechanical Behavior of Materials, Center for Spintronics and Quantum System, School of Materials Science and Engineering, Xi'an Jiaotong University, Xi'an, Shaanxi, 710049, China}
\address{Key Laboratory for Computational Physical Sciences (Ministry of Eduction), Fudan University, Shanghai, 200433, China}

\date{\today}

\begin{abstract}
We propose a new family of the d$^0$ orbital XY (X = K, Rb, Cs; Y = N, P, As, Sb, Bi) monolayers with abundant and novel topology and valley properties. The KN, RbN, RbP, RbAs, CsP, CsAs, and CsSb monolayers possess remarkable quantum anomalous Hall effect (QAHE). CsSb monolayer also exhibits extraordinary valley QAHE with giant splitting. Moreover,  the topological properties of XY monolayers can be efficiently tuned by the in-plane strain, owing to the strain-induced band inversion between the p$_{x,y}$ and p$_z$ orbitals. Our findings suggest that the d$^0$ orbital XY monolayers can be good candidates for promising applications in the spintronics and multifunctional topological-based devices.
\end{abstract}

\maketitle
\section{Introduction}
The quantum anomalous Hall effect (QAHE) is described as a novel quantum state which possesses a finite Chern number \cite{1} and chiral edge states within the insulating bulk \cite{2}. The chiral edge states are topologically guarded and robust against scattering, which provide great potential applications for designing spintronic devices with the low energy consumption and dissipation \cite{3}. The realization of the QAHE needs to break the time reversal symmetry $ \hat{T} $ through magnetism \cite{4,5}. According to the Hund's rule,  spin states are mainly guaranteed by the magnetic elements with partially filled 3d or 4f subshells. Hence, a straightforward way to distinguish quantum anomalous Hall insulator (QAHI) is to screen potential two-dimensional (2D) compounds including d/f block transition metals. Until now, transition metal oxide \cite{6,7,8,9}, transition metal halogenides \cite{10,11,12,13}, and metal organic frameworks (MOF) \cite{14,15,16} have been theoretically predicted. Notably, the QAHE has been experimentally observed in the V/Cr doped (Bi,Sb)$_2$Te$_3$ film \cite{17,18,19} and van der Waals (vdW) layered material MnBi$_2$Te$_4$ \cite{20}, although the QAHE temperature is still very low.

Apart from this class of the d/f orbital QAHI, the QAHE has also been observed in many materials without transition metal, actinide, or rare earth elements,  that is, d$^0$ orbital QAHI. There have been several efforts on obtaining the d$^0$ orbital QAHI, such as moire pattern graphene \cite{21,22}, moire pattern MoTe$_2$/WSe$_2$ \cite{23,24}, silicene/SiC(0001) \cite{25}, stanene/SiC(0001) \cite{26}, and BaSn \cite{3}. The basic design principle of these QAHI to make the p orbital partially occupied, making it have characteristic similar to that of d/f orbital. However, such intrinsic partially occupied p state is difficult to realize in experiments. Exploring new QAHI materials with intrinsic d$^0$-orbital magnetism is a promising way to realize the room-temperature QAHE.

In addition to the QAHE that uses the spin freedom of electron, valley is a new degree of freedom in controlling  electronic transport property \cite{27,28,29,30}, which also has great potential applications in the next generation electronic devices due to its high storage density and low energy consumption \cite{31}. However, the usual 2D materials do not have spontaneous valley splitting on account of the spatial inversion symmetry ($\hat{P}$) \cite{32,33} or the time reversal symmetry $ \hat{T} $ protection. Normally, external strategies, including in use of external magnetic field \cite{35,36}, magnetic proximity effect \cite{37}, or magnetic doping \cite{38}, have to be used to control the valley electronic transport property.  However, these extrinsic methods are difficult to implement for practical applications. Recently, the discovery of 2D ferrovalley (Combined with valley and magnetism freedom) materials has provided a new opportunity to overcome these challenges \cite{39}, although only very few ferrovalley materials have been proposed hitherto \cite{40, 41, 42, 43}. Accordingly, a very important and interesting topic now is to combine the QAHE and valley effect (denoted as VQAHE) in a material. However, due to the stringent criteria of both effects in the real world, most investigations are still limited to the toy model \cite{44,45}.

Based on the first-principles calculations, here we propose a d$^0$-orbital strategy to realize both the intrinsic QAHE and VQAHE in a 2D material, that is, honeycomb lattice XY (X = K, Rb, Cs; Y = N, P, As, Sb, Bi) monolayer. A common feature for these monolayers (except for CsN) is the existence of robust ferromagnetism with the quadratic non-Dirac point. The electronic band structure and Berry curvature calculations show that  KN, RbN, RbP, RbAs, CsP, CsAs, and CsSb monolayers possess remarkable the QAHE, among which the CsSb monolayer is also a new type of ferrovalley material. Moreover,  it is found that the topological properties of XY monolayers can be efficiently tuned by the in-plane strain.

\section{STRUCTURES AND COMPUTATIONAL METHODS}
First, we used the Device Studio to build structure \cite{xx}. Then, to explore the electronic and magnetic structures, we performed the Vienna $Ab$ $initio$ Simulation Package (VASP) \cite{46,47} within the framework of the density functional theory (DFT) for the first-principles calculations. The exchange correlation energy was described by the generalized gradient approximation (GGA) with the Perdew-Burke-Ernzerhof (PBE) functional \cite{48}. The plane-wave basis with a kinetic energy cutoff of 500 eV was employed. $15\times 15\times 1$ and $24\times 24\times 1$ $\Gamma$-centered $k$ meshes were adopted for the structural optimization and the self-consistent calculations. A vacuum of 20 $\rm \AA$ was set along the c-axis, in order to avoid the interaction between the sheet and its periodic images. The total energy convergence criterion and the force were set to be 10$^{-6}$ eV and -0.01 eV/$\rm \AA$, respectively. To investigate the dynamical stability, the phonon spectra were calculated using a finite displacement approach as implemented in the PHONOPY code \cite{49}. The maximally localized Wannier functions (MLWFs) was employed to construct an effective tight-binding Hamiltonian to explore the edge states \cite{50}. Therefore, the edge states were calculated in a half-infinite boundary condition using the iterative Green's function method by the package WANNIERTOOLS \cite{51}. Based on the Heisenberg model, we evaluated the Curie temperature of the XY monolayer systems by using Monte Carlo (MC) simulations.

\section{RESULTS AND DISCUSSION }	
\subsection{Structure and stability}
First of all, we discuss the stability of XY (X = K, Rb, Cs; Y = N, P, As, Sb, Bi) monolayer structure, which presents a honeycomb lattice with the space group $\emph{P-6m2}$ (No. 187) and point group $\emph{D$_{3h}$}$ [Fig. 1 (a-b)]. The primitive unit cell consists of one X and one Y atoms, and the geometry is entirely flat as that of graphene. The relaxed equilibrium lattice constants of XY monolayers are shown in Table I, which obviously increase with the atomic number of X and Y elements. The thermal dynamic stability of XY monolayer is further verified by calculating the phonon dispersion spectrum. As shown in Fig. 2, the absence of imaginary modes (or there is only small imaginary frequencies near $\Gamma$ point) along the high-symmetry lines verifies the dynamical stability of XY monolayers. Note that the small imaginary frequencies near $\Gamma$ point are due to the numerical errors since these imaginary frequencies become smaller as we increase the supercell in the phonon calculations. It is noticed that the 2D room-temperature ferromagnetic (FM) CaCl monolayer has been experimentally synthesized on reduced graphene oxide membranes \cite{52}, which confirms the possibility of synthesizing XY monolayers in experiments.

\begin{table}[htbp]
\caption{
The calculated lattice constants a (\AA) for the monolayer, magnetic ground state (GS), magnetocrystalline anisotropy energy (MAE) (meV/cell), the nearest neighbor exchange interactions parameter J$_0$ (meV), Curie temperature {$\rm T_C$} (K), global band gap E$_g$ (meV), and whether it is a QAHI of the 2D material XY (X = K, Rb, Cs; Y =  N, P, As, Sb, Bi).}
\begin{tabular}{cccccccc}
	\hline
	& a       & GS   & MAE     & J$_0$   & {$\rm T_C$}          & E$_g$   & QAHI   \\
	\hline
	KN      & 4.73    & FM   & 0.07    & 19.00   & 244   & 10.7      & T   \\
	KP      & 5.57    & FM   & 0.37    & 17.52   & 226   & \rule[3pt]{0.2cm}{0.2mm}      & F   \\
	KAs     & 5.72    & FM   & 8.73    & 18.24   & 280   &  \rule[3pt]{0.2cm}{0.2mm}    & F   \\
	KSb     & 6.07    & FM   & 32.30   & 18.15   & 352   &  \rule[3pt]{0.2cm}{0.2mm}    & F   \\
	KBi     & 6.18    & FM   & 72.01   & 18.70   & 436   &  \rule[3pt]{0.2cm}{0.2mm}    & F   \\
	RbN     & 4.96    & FM   & 0.09    & 18.12   & 229   &  16.5     & T   \\
	RbP     & 5.81    & FM   & 0.64    & 15.75   & 196   &  33.2     & T   \\
	RbAs    & 5.95    & FM   & 10.29   & 16.18   & 259   &  26.0     & T   \\
	RbSb    & 6.32    & FM   & 36.23   & 15.62   & 322   &  \rule[3pt]{0.2cm}{0.2mm}    & F   \\
	RbBi    & 6.43    & FM   & 58.92   & 16.28   & 373   &  58.1     & F   \\
	CsN     & 5.15    & AFM   & \rule[3pt]{0.2cm}{0.2mm}    & \rule[3pt]{0.2cm}{0.2mm}   & \rule[3pt]{0.2cm}{0.2mm}   &  \rule[3pt]{0.2cm}{0.2mm}    & F   \\
	CsP     & 6.04    & FM   & 0.25    & 13.10   & 157   &  43.7     & T   \\
	CsAs    & 6.20    & FM   & 10.00   & 13.34   & 220   &  152.0    & T   \\
	CsSb    & 6.58    & FM   & 35.14   & 12.97   & 277   &  154.9    & T   \\
	CsBi    & 6.69    & FM   & 39.18   & 13.45   & 289   &  80.8     & F   \\
	\hline
\end{tabular}
\end{table}	

\subsection{Magnetic property}
An interesting discovery is that all the XY monolayers are magnetic materials. To explore the ground phase of them, the total energies of their nonmagnetic (NM), FM and antiferromagnetic (AFM) states in a 2$\times$2$\times$1 supercell are calculated. Specifically, the ground state of XY monolayers is marked in Table I. It is found that all the XY monolayers have FM state, except for the CsN with a AFM ground state. Our detailed analysis from the computational date shows that these FM monolayers have an integer magnetic moment of 2 $\mu$B per unit cell, which mainly attributed to the $\emph{p}$ orbitals of Y (Y =  N, P, As, Sb, Bi) atoms.

In order to determine the easy magnetization direction of  XY monolayers, we further calculated the magnetocrystalline anisotropy energies (MAE), which is defined as the energy difference between in-plane (100) and out-of-plane (001) spin orientations, i.e., $\Delta E$ = E$_{100}$ - E$_{001}$. The positive value of MAE means that the easy axis is along the z axis. As shown in Table I, for all the monolayers the out-of-plane magnetization direction is more energetically preferred than the in-plane one. Particularly, the MAE of KBi reaches up to 72.01 meV, significantly larger than that of general d/f orbital 2D FM materials \cite{8,11,53,LF}. The large MAE also corresponds to a high Curie temperature from the viewpoint of mean filed theory.

We further calculated the Curie temperature of  XY monolayers based on a 2D Heisenberg model and the Monte Carlo simulations. The Hamiltonian of the 2D Heisenberg model is
\begin{equation}
	H= -\sum_{\langle i,j \rangle}J_0S_i\cdot S_j-\sum_{i} A{S_i}^2,
\end{equation}
where $\emph{J}_0$ and $\emph{A}$ describe the strength of isotropic symmetric spin exchange interactions and single ion anisotropy, respectively. $\langle \emph{i,j} \rangle$ denotes the summation over the nearest neighboring sites, and $\emph{S}$ is the spin moment per atom, respectively.  With the first nearest site pairs merely taken into account, we calculated the exchange interaction parameter $\emph{J}_0$=($E_{AFM}$ - $E_{FM}$)/16$S^2$, where $E_{AFM}$ and $E_{FM}$ are the total energy of the AFM and FM states in a 2$\times$2$\times$1 supercell, respectively. Then, one can obtain $\emph{J}_0$ for XY monolayers, with positive value referring to a FM coupling  (see Table I). Finally, the Monte Carlo simulations were implemented on a 80$\times$80 lattice grid lasting for 100000 steps with a step size of 3 K for the calculation of Curie temperature. As shown in Table I, there are several XY monolayers with  Curie temperature exceeding room temperature (300 K), which may have great potential applications in 2D spintronics.

\subsection{Electronic band structure}
In addition, we explored the spin band structures of XY monolayers. We first show in Fig. 3 the energy bands of XY monolayers without the spin-orbital coupling (SOC) effect. It is seen that the energy bands are fully spin-polarized in these XY monolayers,  i.e., the majority spin bands are insulating while the minority spin bands are metallic. Interestingly, the minority spin bands are  quadratic non-Dirac bands at the high symmetry $\Gamma$ point near the Fermi energy, which may induce the non-trivial topological properties.

So we further calculated the energy bands with SOC effect. As shown in Fig. 4, the energy bands can be segmented into four categories: (1) the quadratic non-Dirac points of the minority spin bands open a gap ranging from 10.7 meV to 152.0 meV at $\Gamma$ point (e.g., KN, RbN, RbP, CsP, CsAs); (2) a direct gap appears in CsBi (80.8 meV) and CsSb (154.9 meV) at $K'$ point; (3) an indirect band gap appears in RbAs (26.0 meV) and RbBi  (58.1 meV); (4) a series of materials are still metallic (e.g., KP, KAs, KSb, KBi, RbSb, CsN). Note that the diverse electronic band structures with SOC band gaps usually forebode the occurrence of abundant topological properties.

\subsection{Quantum anomalous Hall effect}
To reveal the topological properties of XY monolayers, we have calculated the local density of states of the edge state through the Green's function method \cite{54}. As shown in Fig. 5, a single topologically protected edge state appears in between the valence and conduction bands for all the XY monolayers, except for KSb, KBi, RbBi and CsBi. To confirm the topological characteristic and evaluate the  anomalous Hall conductivity (AHC),  we also performed the Berry curvature calculations using the formula

\begin{equation}
	\sigma_{xy} = C\frac{e^2}{h},
\end{equation}
\begin{equation}
	C= \frac{1}{2\pi} \int_{BZ} d^2k ~\Omega(\textbf{k}),
\end{equation}

\begin{equation}
	\Omega(\textbf{k})=-\sum_{n}f_{n}\sum_{n\prime \neq n}\frac{2Im \left \langle \psi_{nk} \mid v_{x} \mid \psi_{n\prime k} \right \rangle \left \langle \psi_{n\prime k} \mid v_{y} \mid \psi_{nk} \right \rangle}{(E_{n\prime}-E_{n})^2},
\end{equation}
where C is Chern number, $\Omega(\textbf{k})$ is the Berry curvature in the reciprocal space, $v_{x}$ and $v_{y}$ are operator components along the x and y directions and $f_{n}=1$ for the occupied bands, respectively \cite{55,56,57}. One can obtain the Chern number as well as AHC by integrating the Berry curvature in the entire BZ.

For instance, we show in Fig. 6(c) the calculated Berry curvature distribution of CsSb monolayer, the integration of which gives rise to the Chern number C=1, confirming the nontrivial topological property. Moreover, the calculated energy-level dependence of AHC shows that a quantized platform appears at the Fermi level [Fig. 6(e)], confirming the QAHI characteristic of CsSb. The topological characteristic for other XY monolayers are additionally shown in Table. I, where 7 of 15 XY monolayers present the QAHI characteristic. Particularly, the topological band gaps of CsAs and CsSb are even larger than 150 meV, showing that they can be ideal candidates for the high-temperature QAHI devices. The above results show that the 2D honeycomb lattice composed of binary anions and cations may hold abundant anomalous topological properties, which are deserved for further investigations.

\subsection{Valley quantum anomalous Hall effect}

In addition to the amazing QAHE discussed above, the XY monolayers can also present exotic ferrovalley property. Fig. 6(a) and (b) show the spin band structures of CsSb under different magnetization directions with the SOC effect, which display the remarkable valley futures at K and K' points. This is because the broken time-reversal, spatial inversion symmetry and the SOC effect collectively break the degeneracy between the K and K' valley states for valence and conduction bands. The valley polarization appears at both valence and conduction bands. Moreover, the energy of K valley state is lower than that of K' in the valence band, which results in a spontaneous valley polarization with a large valley splitting of 257.8 meV. Whereas, the energy of K valley state is higher than that of K' in the conduction band, making a spontaneous valley polarization with a valley splitting of 223.6 meV. As a result, the total valley splitting gap becomes 481.4 meV in the CsSb monolayer, which is much larger than that in other ferrovalley materials, e.g., Nb$_{3}$I$_{8}$(107 meV) \cite{41}, TiVI$_6$(22 meV) \cite{42} and LaBr$_2$ (33 meV) \cite{43}. This huge gap value shows the extreme robustness of valley polarization against the external perturbations. It is noticed that, when the magnetization direction of  CsSb is reversed, its valley polarization reverses accordingly. Therefore,  the spin and valley polarizations will be simultaneously flipped by reversing the magnetization direction [Fig. 6(b)]. This feature indicates that manipulating the magnetization direction can be an effective way to alter the valley states.

In order to understand the underlying mechanism for the ferrovalley effect in CsSb, we tried to construct an effective Hamiltonian model based on the DFT calculations. As shown in Fig. S1, the valence band maximum (VBM) and conduction band minimum (CBM) are mainly composed of Sb p$_x$ and p$_y$ orbitals under the SOC effect. Considering that the little group of K and K' points belongs to D$_{3h}$ in the out-of-plane magnetization case, we adopted $|$$\psi$$_v$$^{\tau}$$\rangle$=$\frac{1}{\sqrt{2}}$($|$p$_x$$\rangle$+i$\tau$$|$p$_y$$\rangle$)$\otimes$$|$$\downarrow$$\rangle$, $|$$\psi$$_c$$^{\tau}$$\rangle$=$\frac{1}{\sqrt{2}}$($|$p$_y$$\rangle$+i$\tau$$|$p$_x$$\rangle$)$\otimes$$|$$\downarrow$$\rangle$ as the orbital basis for the VBM and CBM, where $\tau$ = $\pm$1 indicate the valley index corresponding to K/K'. Since the VBM and CBM belong to the same spin channel (spin down bands), we take the SOC effect as the  perturbation term, which is
\begin{equation}
\hat{H}_{SOC} = \lambda \hat{S} \cdot \hat{L} = \hat{H}_{SOC}^{0} + \hat{H}_{SOC}^{1},
\end{equation}
where $\hat{S}$ and $\hat{L}$ are spin angular and orbital angular operators, respectively. $\hat{H}_{SOC}^{0}$ and $\hat{H}_{SOC}^{1}$ represent the interaction between the same spin states and between opposite spin sates, respectively. For the CsSb monolayer, the single valley is composed of only one spin channel [see Fig. 3(l)], and the other spin channel is far from the valleys. Hence, the term $\hat{H}_{SOC}^{1}$ can be ignored. On the other hand,  $\hat{H}_{SOC}^{0}$ can be written in polar angles
\begin{equation}
\hat{H}_{SOC}^{0} = \lambda \hat{S}_{z'}(\hat{L}_zcos\theta + \frac{1}{2}\hat{L}_+e^{-i\phi}sin\theta + \frac{1}{2}\hat{L}_-e^{+i\phi}sin\theta),
\end{equation}
In the out-of-plane magnetization case, $\theta$ = $\phi$ = 0, then the $\hat{H}_{SOC}^{0}$ term can be simplified as
\begin{equation}
\hat{H}_{SOC}^{0} = \lambda \hat{S}_{z} \hat{L}_z,
\end{equation}
The energy levels of the valleys for the VBM and CBM can be expressed as E$_v$$^ \tau$ = $\langle$ $\psi$$_v$$^ \tau$ $|$ $\hat{H}$$_{SOC}^{0}$ $|$ $\psi$$_v$$^ \tau$ $\rangle$ and E$_c$$^ \tau$ = $\langle$ $\psi$$_c$$^ \tau$ $|$ $\hat{H}$$_{SOC}^{0}$ $|$ $\psi$$_c$$^ \tau$ $\rangle$, respectively. Then, the valley polarization in the valence and conduction bands can be expressed as
\begin{equation}
E_{v}^{K} - E_{v}^{K'} = i \langle p_x | \hat{H}_{SOC}^{0} | p_y \rangle - i \langle p_y | \hat{H}_{SOC}^{0} | p_x \rangle \approx \lambda,
\end{equation}
\begin{equation}
E_{c}^{K} - E_{c}^{K'} = i \langle p_y | \hat{H}_{SOC}^{0} | p_x \rangle - i \langle p_x | \hat{H}_{SOC}^{0} | p_y \rangle \approx \lambda,
\end{equation}
where the $\hat{L}_z|p_x \rangle$ = i$\hbar$$|p_y \rangle$, $\hat{L}_z|p_y \rangle$ = -i$\hbar$$|p_x \rangle$. The analytical result certificates that the valley degeneracy splits for the valence and conduction bands are consistent with our DFT calculations ($E_{v}^{K'}$ - $E_{v}^{K}$ = 257.8 meV, $E_{c}^{K}$ - $E_{c}^{K'}$ = 223.6 meV).

It should be noted that the valley polarization in CsSb monolayer is different from that previously reported \cite{39,40,41,43}, whose split only occurs at the VBM ($E_{v}^{K}$ $\neq$ $E_{v}^{K'}$, $E_{c}^{K}$ = $E_{c}^{K'}$) or CBM ($E_{v}^{K}$ = $E_{v}^{K'}$, $E_{c}^{K}$ $\neq$ $E_{c}^{K'}$). Due to the strong SOC effect in the CsSb monolayer, the valence (conduction) bands moves upwards (downwards) at the K' valley [relative to no SOC band structure, Fig. 3(l)], while the valence (conduction) bands shifts downwards (upwards) at the K valley. Thus, the valley degeneracy is broken at both K and K' point, which induces the significant valley polarizations. As a result, the novel phenomenon appears, i.e., the K valley is annihilated and the K' valley remains (we denote it as single valley polarization).

The above results show that CsSb monolayer holds both remarkable QAHE and ferrovalley properties,  indicating it is an realistic VQAHE material beyond the toy model. To ensure this,  we have calculated the Berry curvature of CsSb monolayer at K and K' points, respectively. As shown in Fig. 6 (c) and (d), the  Berry curvatures at  K and K' points have opposite signs, showing the typical valley polarization characteristic. By integrating the Berry curvature over the BZ, one can further calculate the AHC. As shown in Fig. 6 (e), a valley-polarized Hall conductivity clearly exists in the CsSb monolayer. Specifically, when the Fermi level lies between the VBM or CBM of the K and K' valleys, the valley-polarized Hall conductance ${\sigma}_{xy}$ can be achieved. This result confirms  the existence of VQAHE in the CsSb monolayer. Moreover, in the hole doping condition,  when the magnetism direction of  CsSb is in +z direction,  the spin-up holes from the K' valley will be generated and accumulate on  one boundary of the sample under an in-plane electrical field [upper plane of Fig. 6(f)].  On the other hand, when the magnetism direction is in -z direction,  the spin-up holes from the K valley will be generated and accumulate on the opposite boundary of the sample under an in-plane electrical field [lower plane of Fig. 6(f)]. This feature shows that monolayer CsSb is an ideal candidate for the high-performance valleytronic devices.

\subsection{Strain-induced topological insulator phase transitions}

It is noticed that an interesting  topological phase transition from QAHI to NI appears with the atomic number of Y element increasing  for the magnetic XY monolayers (see Table. I). For instance,  in  CsY (Y = P, As, Sb, Bi), the topological phase transition occurs between CsSb (QAHI) and CsBi (NI). Through the detailed analysis on the energy band variation with  atomic number of Y element, it is found that the topological properties are closely related to the gap at K' point. As shown in Fig. 7(a), with the atomic number of Y increases, the gap at K' point gradually decreases to zero in CsY. The fitting curve shows that the band gap would  reach zero when Y is in between  Sb and Bi, which is the phase transition point. We further analyzed the energy level assignments of Sb-p and Bi-p orbitals. As shown in Fig 7(b), there is a band inversion between p$_{x,y}$ and p$_z$ orbitals in CsSb, which is a characteristic of QAHI. Whereas, the band inversion disappears in CsBi due to the stronger of SOC effect [Fig. 7(c)], and thus CsBi becomes a NI. Note that this feature is also applicable in KY (Y = N, P, As, Sb, Bi) and RbY (Y = N, P, As, Sb, Bi).

Considering that the band gap and SOC in 2D materials are usually sensitive to the lattice strain, we further explored the strain manipulation on the topological phase transition in XY monolayers. In the following, we mainly present the results of CsBi monolayer, which has a NI phase in the strain-free condition.  In the calculations, we adopted the in-plane biaxial strain, which can be defined as $\varepsilon$ = (a-a$_0$)/a$_0$$\times$100$\%$. In the formula, a and a$_0$ represent lattice constant after and before in-plane biaxial strain is applied, respectively. As shown in Fig. 8(a), the band gap decreases first with the increase of compressive strain, which becomes almost zero when the strain reaches 6$\%$. Whereas, it starts to increase when a larger compressive strain is applied, which is expected to induce a transition from NI to QAHI phase. Specifically, we show in Fig. 8(b) and (c) the calculated edge states of CsBi under 0$\%$ and -10$\%$ strains, respectively. One can see that the edge state does not exist under the 0$\%$ strain, while it clearly appears under the -10$\%$ strain, consistent with the phase diagram shown in Fig. 8(a). On the other hand, we failed to observe the phase transition from NI to QAHI when a stretch strain is applied. This result indicates that only the compressive strain can induce the band inversion, and thus leads to the topological phase transition in CsBi monolayer.

To further demonstrate the universality of the strain-induced band inversion mechanism in the XY system, we systematically explored the topological properties of CsSb (Fig. S1, S2), CsP (Fig. S3, S4), CsBi (Fig. S5), and RbBi (Fig. S6, S7) under different strains. It is found that strain can induce the band inversions which lead to the topological phase transition from the NI to QAHI in these systems. This result indicates that the strain-induced band inversion is applicable to the entire XY system. Therefore, our discovery would be beneficial to the experimental exploration for  the high-temperature QAHE and VQAHE in the $d^0$ orbital materials.


\section{CONCLUSION}
In conclusion, by using first-principles calculations, we have proposed a new family of the d$^0$ orbital XY (X = K, Rb, Cs; Y = N, P, As, Sb, Bi) monolayers with abundant and novel topology and valley properties. We have found that the band structures of all XY monolayers possess quadratic non-Dirac point at $\Gamma$ point, and the nontrivial gap in range of 10.7 $ \thicksim $ 154.9 meV can be induced by the considerable SOC effect in this 2D system. Through the Berry curvature and edge state calculations, we have found that KN, RbN, RbP, RbAs, CsP, CsAs, and CsSb monolayers present remarkable QAHE, and CsSb monolayer also exhibits extraordinary VQAHE with giant polarization splitting between K and K' points. Taken the CsBi monolayer as an instance, we have additionally discovered that the topological properties of XY monolayers can be efficiently tuned by the in-plane strain, due to the strain-induced band inversion between the p$_{x,y}$ and p$_z$ orbitals. The abundant topological properties in XY monolayers make them promising candidates for the applications in the spintronic nanodevices.

\section*{ACKNOWLEDGEMENTS}
We thank Prof. Huisheng Zhang for valuable discussions. This work is supported by National Natural Science Foundation of China (No. 12004295 and No. 12074301), the National Key R\&D Program of China (2017YFA0206202). P. Li thanks China's Postdoctoral Science Foundation funded project (No. 2020M673364), the Open Project of the Key Laboratory of Computational Physical Sciences (Ministry of Education), and the Open Project of State Key Laboratory of Surface Physics (No. KF2022$\_$09). K. Wang, Y. H. Li, and H. L. Mei thanks the Undergraduate Innovation and Entrepreneurship Training Program of Shaanxi Province (No. SJ202010698179). We gratefully acknowledge HZWTECH for providing computation facilities, and the computational resources provided by the HPCC platform of Xi'an Jiaotong University.


\begin{figure}[htb]
	\begin{center}
		\includegraphics[angle=0,width=0.8\linewidth]{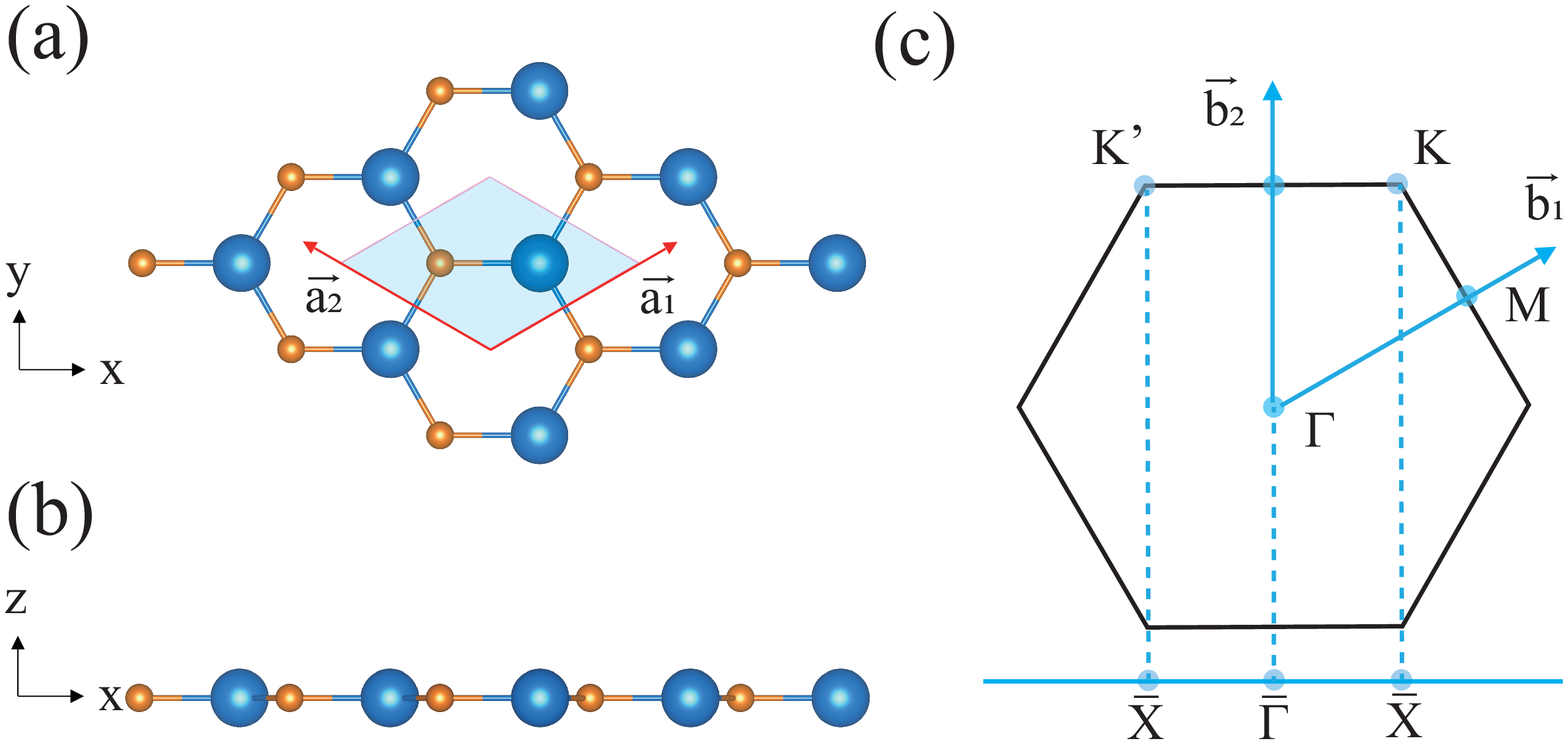}
		\caption{(a) Top and (b) side view of the lattice structure for XY (X = K, Rb, Cs; Y=  N, P, As, Sb, Bi) monolayer with lattice vectors $\vec{a}_1$ and $\vec{a}_2$. The X (X = K, Rb, Cs) and Y (Y = N, P, As, Sb, Bi) atoms are depicted by the blue and orange balls, respectively. The rhombus covered with light blue shadow denotes the unit cell. (c) The Brillouin zone (BZ) of the honeycomb lattice with the reciprocal lattice vectors $\vec{b}_1$ and $\vec{b}_2$. $\Gamma$, K and M are the high-symmetry points in the BZ, and $\overline{\Gamma}$ and $\overline{X}$ are the high-symmetry points in the one-dimensional BZ.}
	\end{center}
\end{figure}

\begin{figure}[htb]
	\begin{center}
		\includegraphics[angle=0,width=0.80\linewidth]{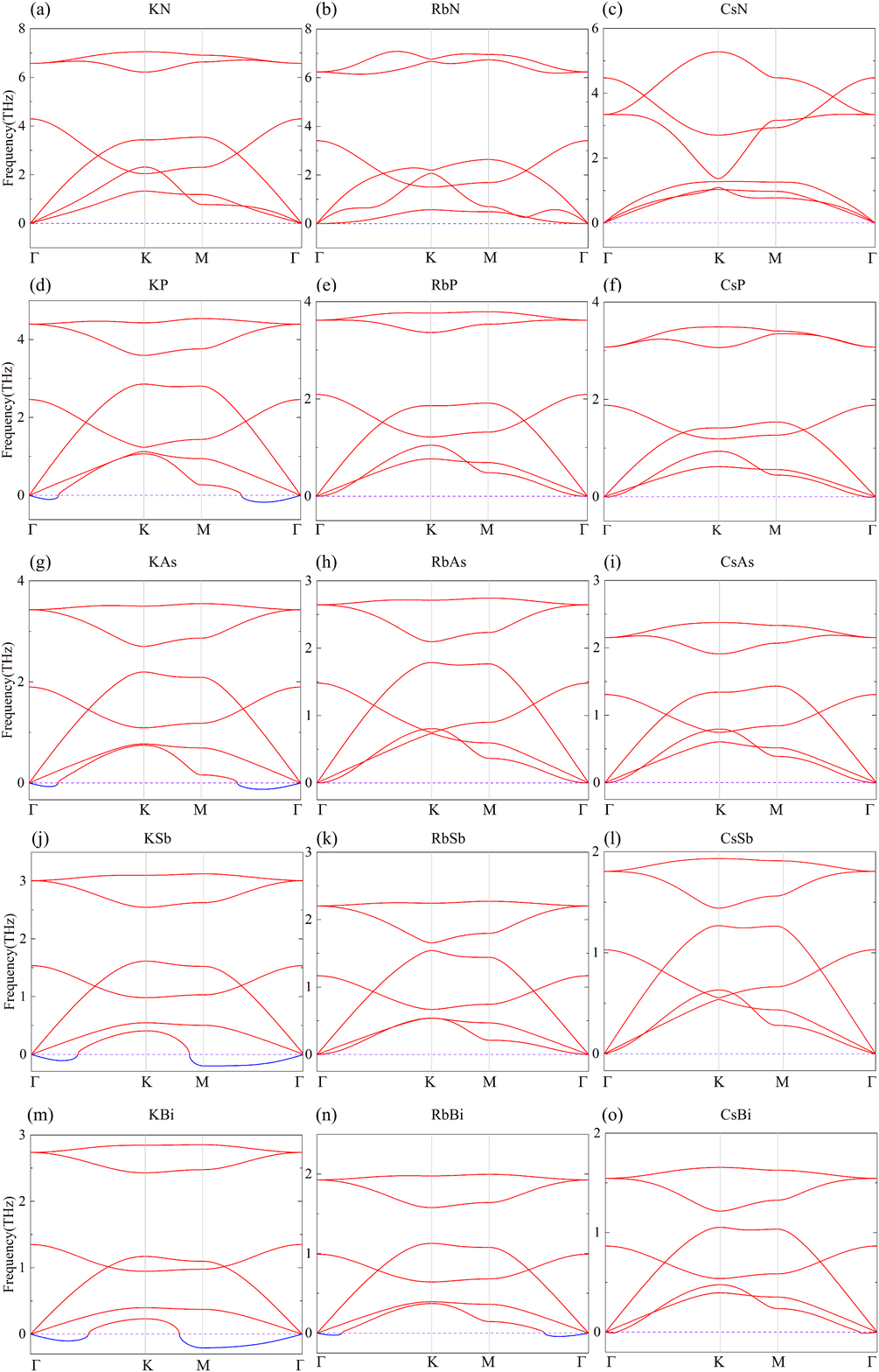}
		\caption{(a-o) The calculated phonon dispersion curves along the high-symmetry lines within the first BZ of the (a) KN, (b) RbN, (c) CsN,  (d) KP, (e) RbP, (f) CsP, (g) KAs, (h) RbAs, (i) CsAs, (j) KSb, (k) RbSb, (l) CsSb, (m) KBi, (n) RbBi and (o) CsBi, respectively. \textcolor{red}{The imaginary frequencies in the phonon dispersion spectrum are marked blue while other parts red.}}
	\end{center}
\end{figure}

\begin{figure}[htb]
	\begin{center}
		\includegraphics[angle=0,width=0.8\linewidth]{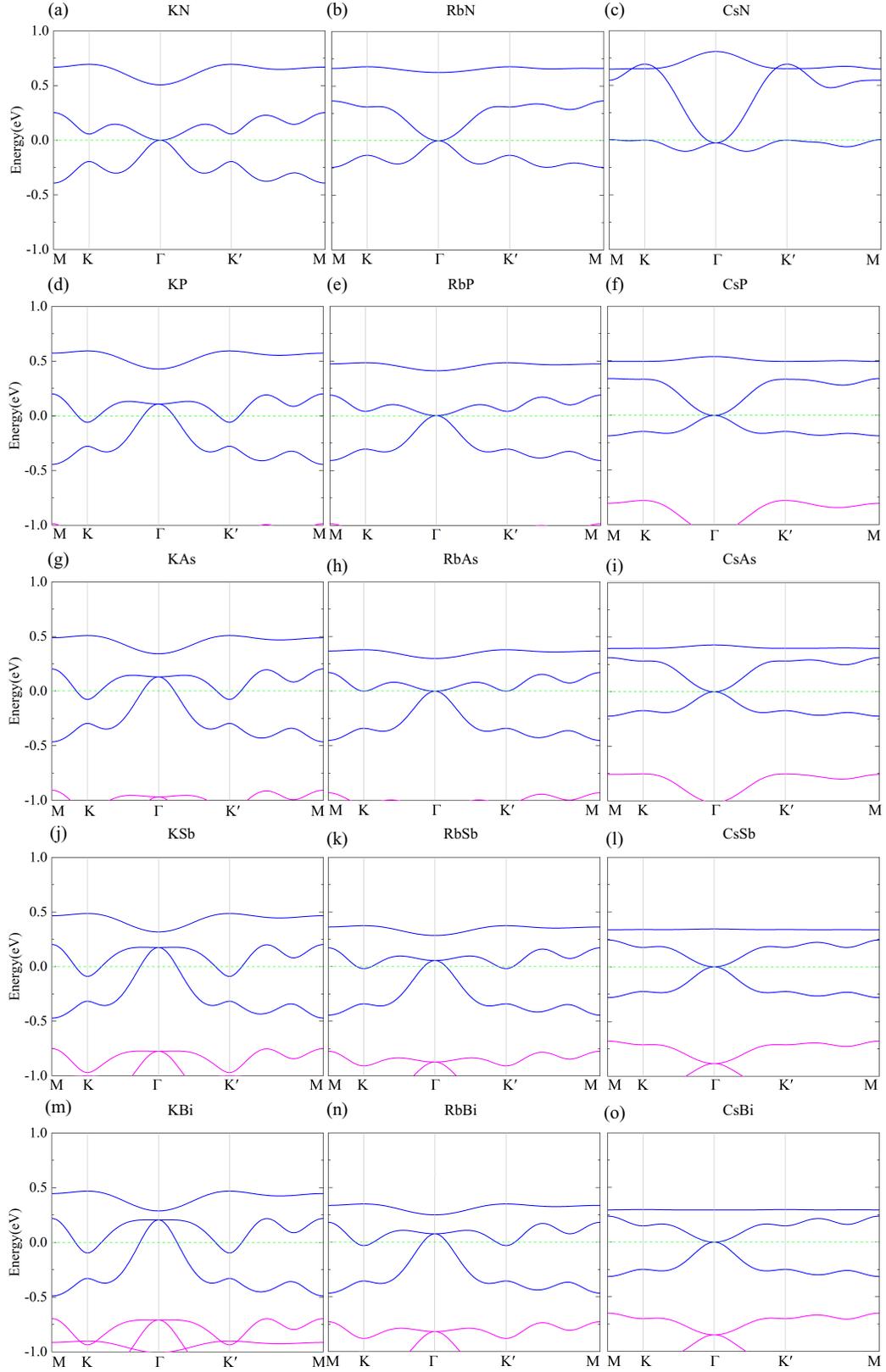}
		\caption{(a-o) Spin-polarized band structures of (a) KN, (b) RbN, (c) CsN,  (d) KP, (e) RbP, (f) CsP, (g) KAs, (h) RbAs, (i) CsAs, (j) KSb, (k) RbSb, (l) CsSb, (m) KBi, (n) RbBi and (o) CsBi, respectively. The magenta, and blue lines represent spin up and spin down bands, respectively.}
	\end{center}
\end{figure}

\begin{figure}[htb]
	\begin{center}
		\includegraphics[angle=0,width=0.8\linewidth]{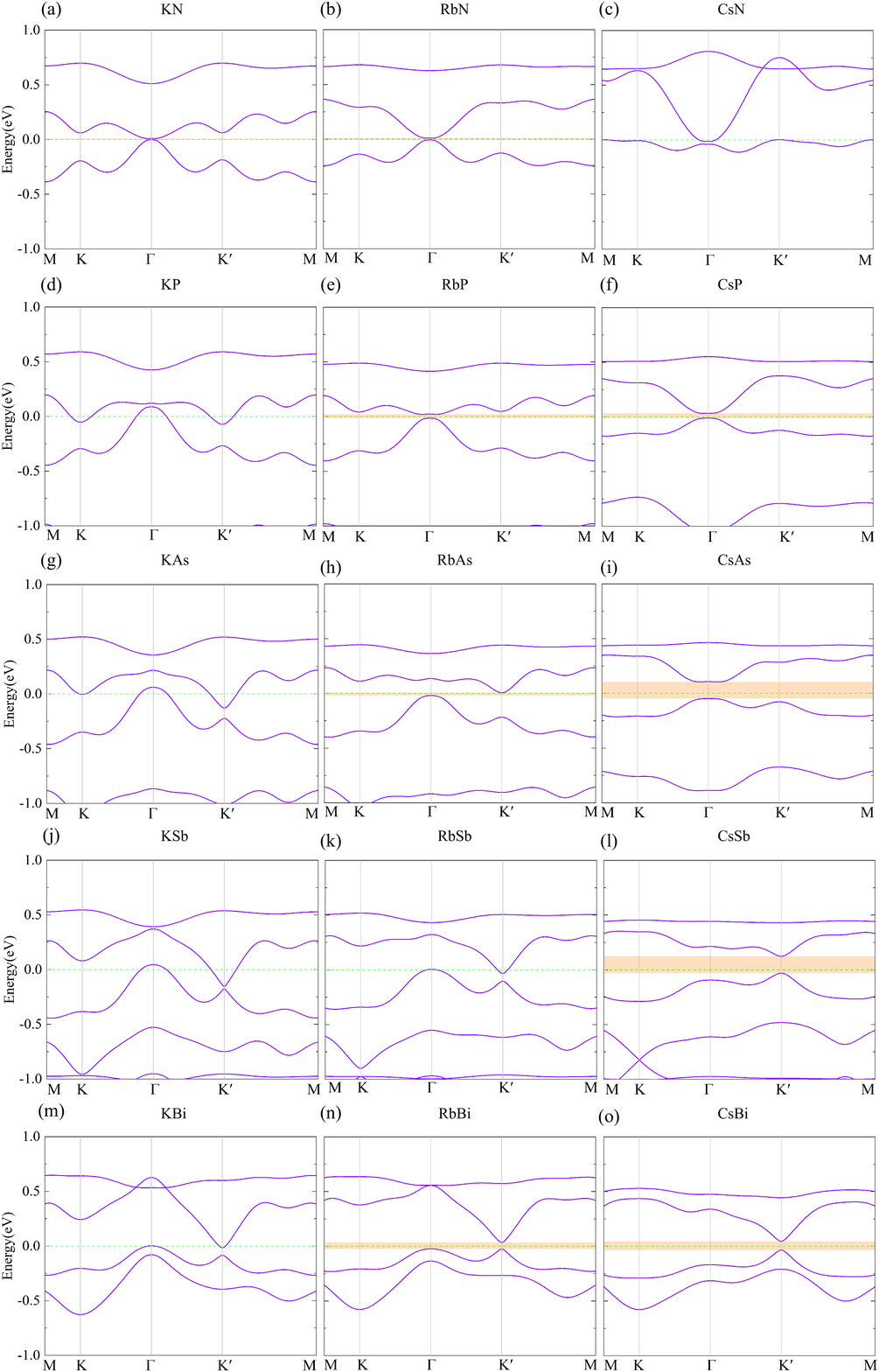}
		\caption{ (a-o) Band structures with SOC of (a) KN, (b) RbN, (c) CsN,  (d) KP, (e) RbP, (f) CsP, (g) KAs, (h) RbAs, (i) CsAs, (j) KSb, (k) RbSb, (l) CsSb, (m) KBi, (n) RbBi and (o) CsBi, respectively. The band gap is indicated by the orange shading. }
	\end{center}
\end{figure}

\begin{figure}[htb]
	\begin{center}
		\includegraphics[angle=0,width=0.8\linewidth]{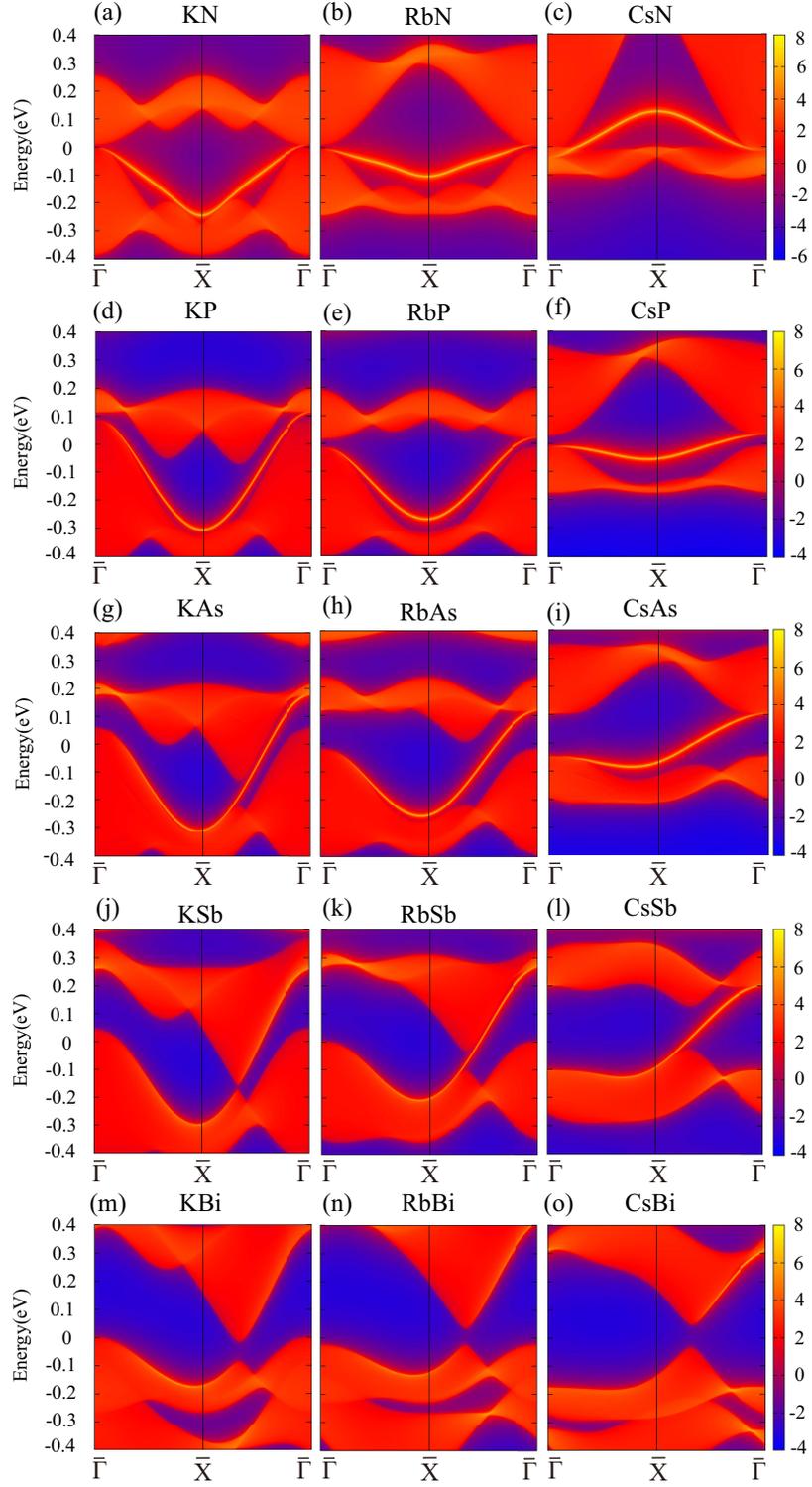}
		\caption{ (a-o) Edge states of the (a) KN, (b) RbN, (c) CsN,  (d) KP, (e) RbP, (f) CsP, (g) KAs, (h) RbAs, (i) CsAs, (j) KSb, (k) RbSb, (l) CsSb, (m) KBi, (n) RbBi and (o) CsBi, respectively.}
	\end{center}
\end{figure}

\begin{figure}[htb]
	\begin{center}
		\includegraphics[angle=0,width=0.8\linewidth]{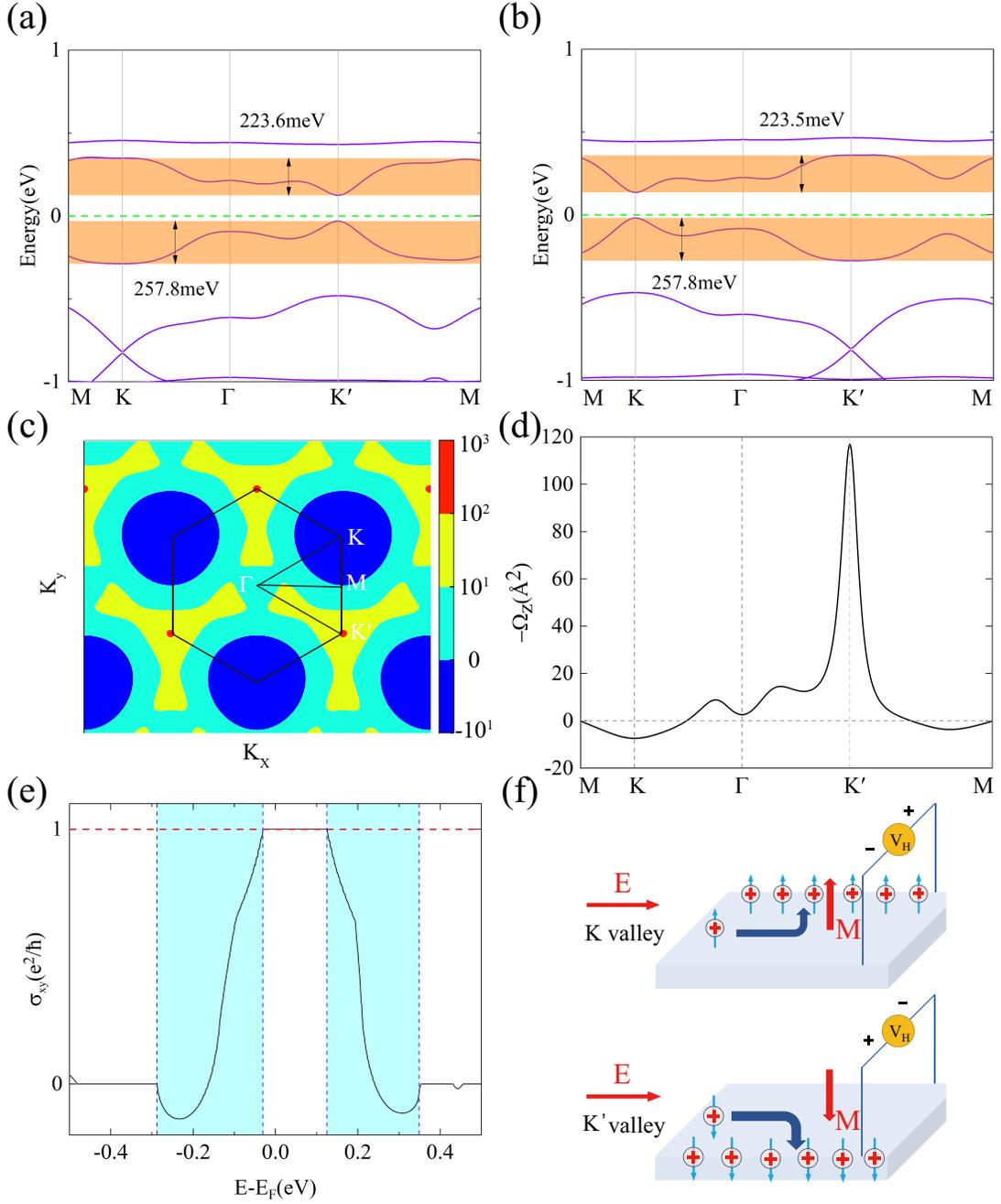}
		\caption{
			The band structure of CsSb with SOC for magnetic moment of Sb along the positive (a) and negative (b) z direction (out of plane), respectively. The Berry curvatures of CsSb in the Brillouin zone (c) and along the high symmetry line (d). (e) Calculated AHC ${\sigma}_{xy}$ as a function of Fermi energy for CsSb monolayer. The two light blue shadows denote the valley splitting between K and K' valley. (f) Schematic diagram of tunable valley quantum anomalous Hall effect in hole doped CsSb monolayer at the K and K' valley, respectively. The holes are denoted by + symbol. Upward arrows and downward arrows refer to the spin-up and spin-down carriers, respectively.  }
	\end{center}
\end{figure}

\begin{figure}[htb]
	\begin{center}
		\includegraphics[angle=0,width=0.80\linewidth]{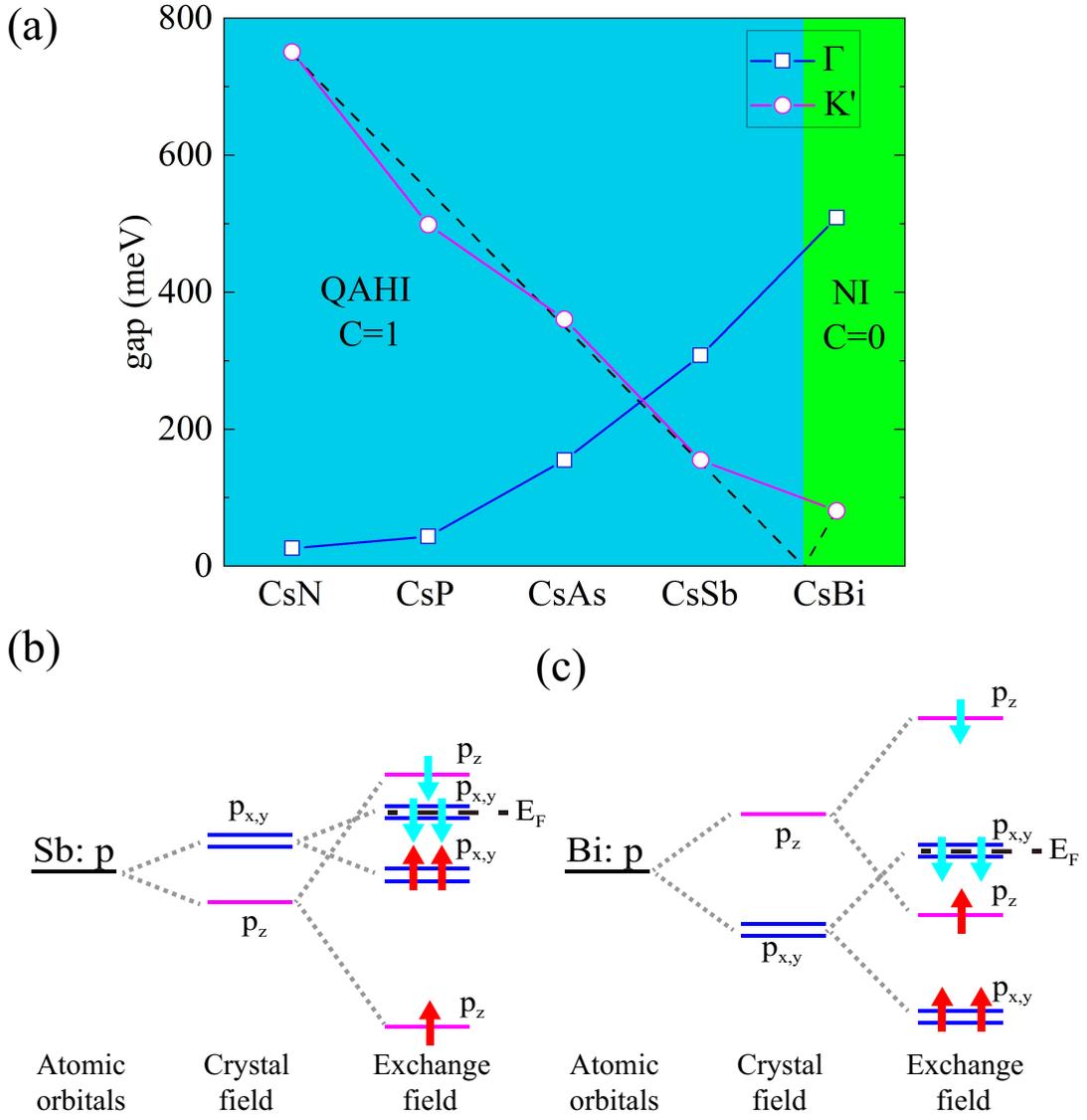}
		\caption{
			(a) The gap of CsY (Y = N, P, As, Sb, Bi) at $\Gamma$ and K' points. The dotted line is the fitting K' point gap variation trend. (b, c) The schematic illustration of energy level for Sb-p orbital in CsSb (b) and Bi-p orbital in CsBi (c). }
	\end{center}
\end{figure}

\begin{figure}[htb]
	\begin{center}
		\includegraphics[angle=0,width=0.8\linewidth]{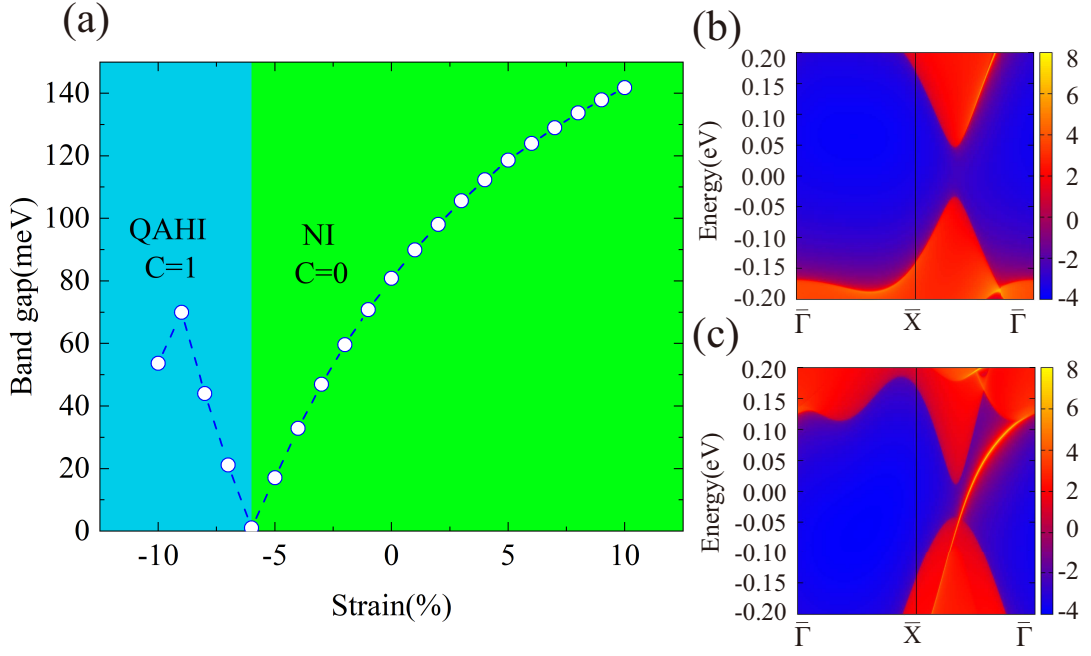}
		\caption{
			 \textcolor{red}{In-plane biaxial strain effect on the phase transition between QAHI and NI in CsBi monolayer.} (a) The band gap as a function of in-plane biaxial strains from -10$\%$ to 10$\%$ , where blue and green regions represent the QAHI and NI, respectively. The negative and positive percentage refer to the compressive and tensile strain, respectively. (b-c) The calculated edge states of a semi-infinite sheet for the under 0$\%$ (b) and -10$\%$ (c) compressive strain.}
	\end{center}
\end{figure}

\end{document}